\documentclass[prb,twocolumn]{revtex4}

\usepackage{graphicx}

\begin{document}

\title{Role of covalent Fe-As bonding in the magnetic moment formation
and exchange mechanisms in iron-pnictide superconductors}

\author{K. D. Belashchenko}
\affiliation{Department of Physics and Astronomy and Nebraska Center
for Materials and Nanoscience, University of Nebraska--Lincoln,
Lincoln, Nebraska 68588, USA}

\author{V. P. Antropov}
\affiliation{Condensed Matter Physics, Ames Laboratory, Ames, Iowa 50011, USA}

\date{\today}

\begin{abstract}
The electronic origin of the huge magnetostructural effect in layered Fe-As
compounds is elucidated using LiFeAs as a prototype. The crucial feature of
these materials is the strong covalent bonding between Fe
and As, which tends to suppress the exchange splitting. The bonding-antibonding
splitting is very sensitive to the distance between Fe and As nuclei. We argue
that the fragile interplay between bonding and magnetism is universal for this
family of compounds. The exchange interaction is analyzed in real space, along
with its correlation with covalency and doping. The range of interaction and
itinerancy increase as the Fe-As distance is decreased. Superexchange makes a
large antiferromagnetic contribution to the nearest-neighbor coupling, which
develops large anisotropy when the local moment is not too small. This
anisotropy is very sensitive to doping.
\end{abstract}

\maketitle

Layered iron-pnictide compounds have recently attracted a lot of interest due
to their high superconducting transition temperature. \cite{Kamihara} Pairing
is widely believed to be mediated by spin fluctuations in these materials;
\cite{Mazin-SC} understanding of magnetic interaction is therefore of utmost
importance. Numerous first-principles calculations revealed a huge
magnetostructural effect manifesting itself, in particular, in the strong
sensitivity of the Fe local moments to the Fe-As distance $R_{\mathrm{Fe-As}}$.
\cite{Yin}

Tight-binding parameterizations of the band structure indicate that Fe-As
hybridization is significant,\cite{Cao,Tesanovic} and it was used to explain
the instability of the Fe local moments\cite{Tesanovic,Wu} (a similar effect is
known for zincblende iron pnictides\cite{Mirbt}). However, \emph{ab initio}
results have been mainly described in terms of weak \cite{Boeri,Yildirim} Fe-As
hybridization comparable to oxides. \cite{Singh-Du} Strong Fe-As mixing was
mentioned,\cite{Yin,Vildosola} but its role in the magnetism was not
explained. Ref.\ \onlinecite{Yildirim} focused on the effect of the local
moment on As-As bonding. In this paper we show that the coupling between the
local moment and the Fe-As distance is controlled by strong covalent Fe-As
bonding, and analyze its effects on the exchange interaction in the (likely
ground-state) ``stripe'' phase using the linear response technique. The salient
features of chemical bonding and its relation to magnetism appear to be
universal across the whole family of iron-pnictide layered materials, and we
chose LiFeAs, which is a superconductor below 18 K, \cite{Wang} as a
representative example.

\begin{figure*}
~\hfill\includegraphics*[width=0.26\textwidth]{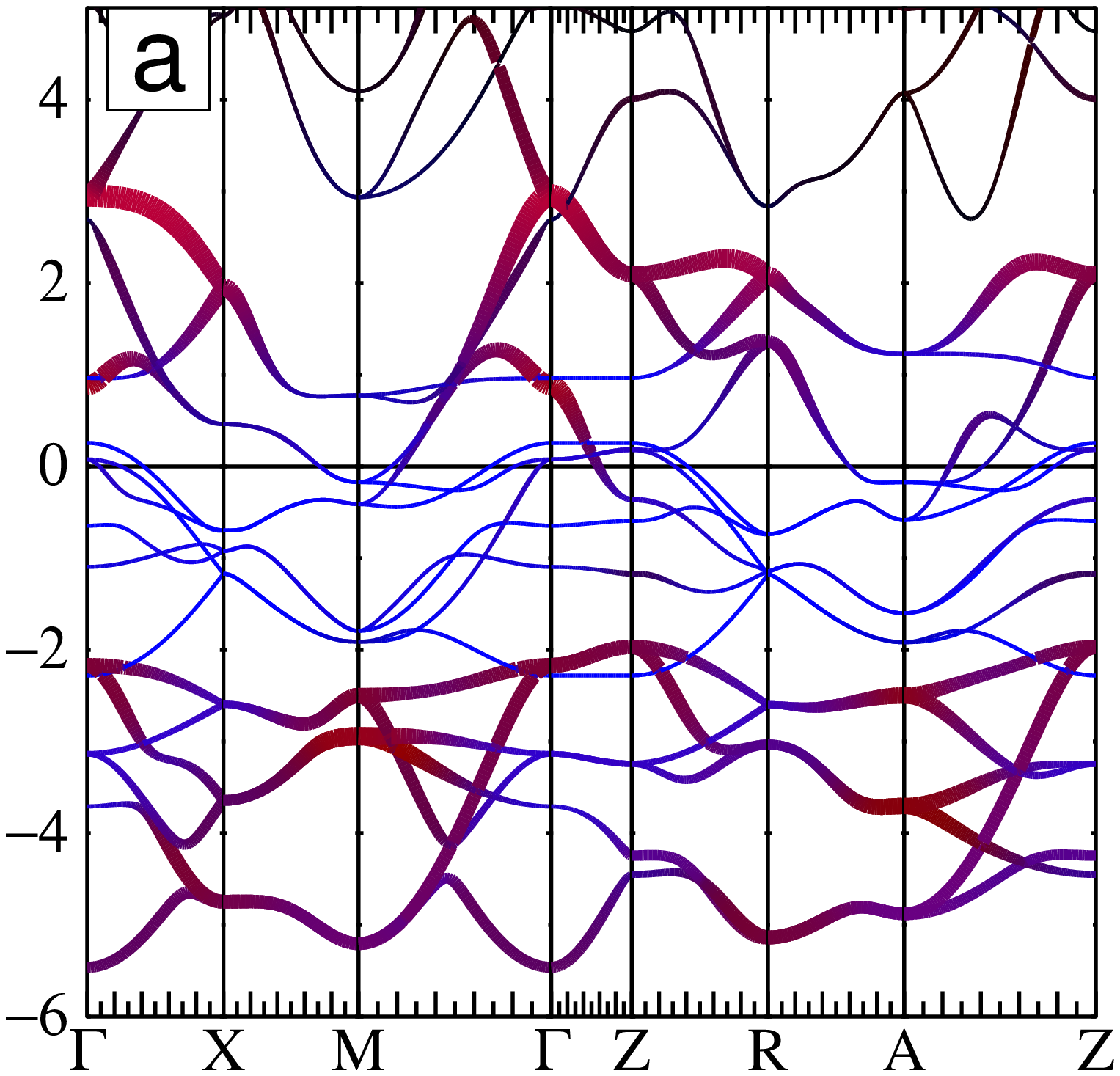}\hfill
\includegraphics*[width=0.26\textwidth]{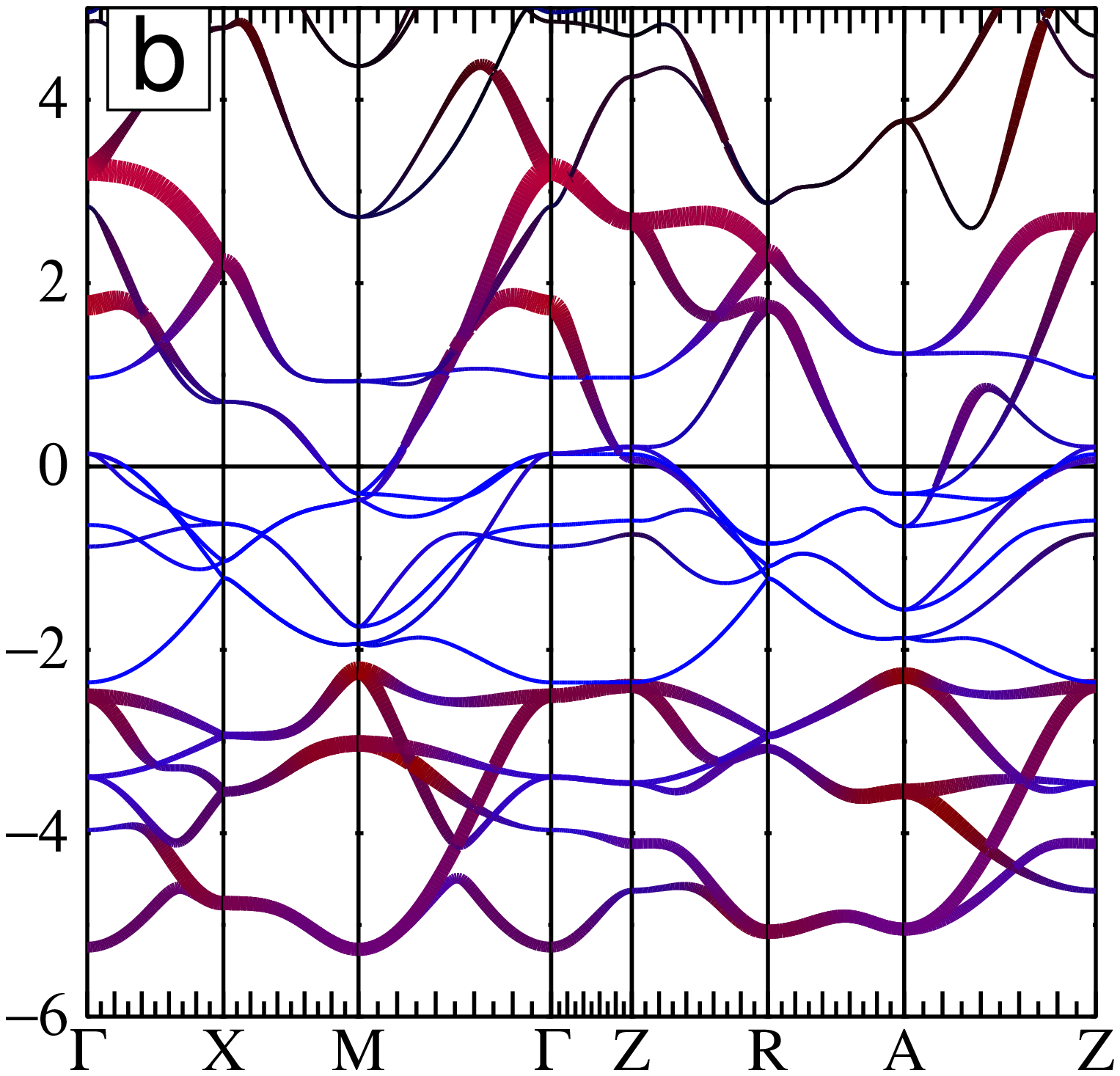}\hfill
\includegraphics*[width=0.26\textwidth]{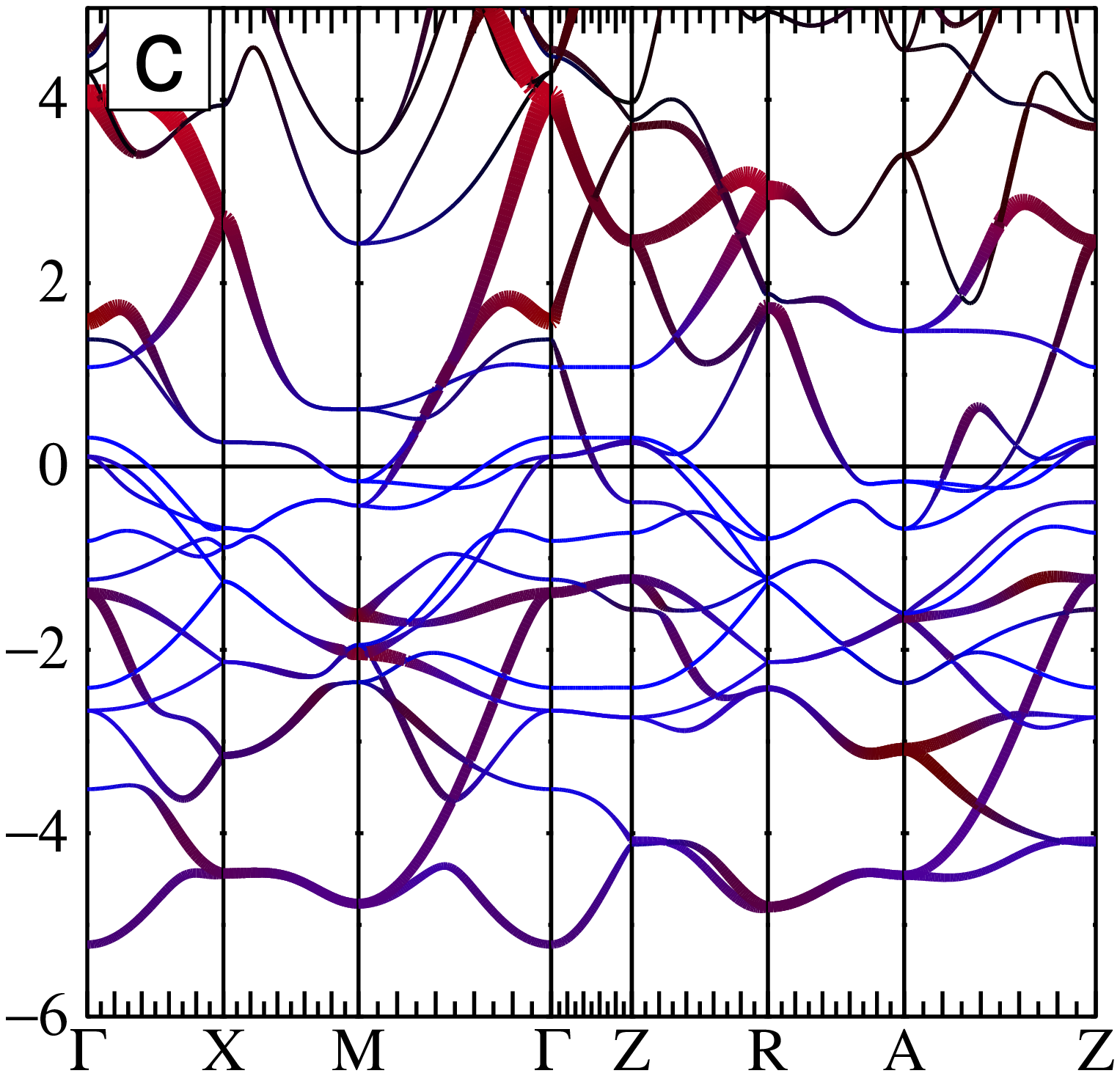}\hfill~
\caption{Energy bands of LiFeAs in the non-magnetic state. Mulliken weights of
As ($4p$) orbitals and Fe ($3d$ and $4s$) orbitals are shown by red and blue
color, respectively. Bands with no contributions from these states appear black;
fully hybridized bands appear purple. In addition, the As ($4p$) weight is
shown by line thickness. (a) Experimental lattice parameters with
$R_{\mathrm{Fe-As}}=2.42$ \AA. (b) $R_{\mathrm{Fe-As}}$ reduced to 2.33 \AA.
(c) External potential of 5 eV applied to the As $p$ states (see text).}
\label{bands}
\end{figure*}

The band structure of LiFeAs was calculated by Singh; \cite{Singh} in most
respects it is similar to LaFeAsO, BaFe$_2$As$_2$, and other members of this
layered iron-pnictide family. Let us analyze the orbital content of the Bloch
states.  Fig.\ \ref{bands}a shows the energy bands \cite{ASA} of non-magnetic
LiFeAs calculated for the experimental structure. \cite{lattice} In this
picture, As ($4p$) weight is shown in red color and also using line thickness;
the iron ($4s$ and $3d$) weight is shown in blue. We have verified that the
weights of different $3d$ cubic harmonics on Fe sites (not shown) are similar
to those plotted in Ref.\ \onlinecite{Boeri} for LaFeAsO. It is seen
from Fig.\ \ref{bands}a that As and Fe states form fully mixed bonding and
antibonding states centered, respectively, at 3.5 eV below and at 1.5 eV above
the Fermi level $E_F$. The large bonding-antibonding splitting of about 5 eV
indicates very strong covalent bonding between As $p$ and Fe $d$ states,
in the sense that the splitting is large compared to the bare level separation.
The hybridized bands have almost equal weights of Fe and As states, and the As
states contribute equally to the bonding and antibonding states. This picture is somewhat
different from that presented for LaFeAsO,\cite{Vildosola} where the bands were divided
in an upper group of mainly Fe $3d$ bands and a lower group of mainly pnictogen and oxygen $p$ bands.
The separation of As states in two subbands
can only be explained by hybridization with Fe. Indeed, the As states of a
fictitious system with Fe atoms removed from the lattice form a gapless set of
bands about 4 eV wide; this was checked using the FLAPW method.\cite{ASA}
We also note that although the bare Fe $4s$ states are a few volts above $E_F$,
they contribute appreciably at 5-6 eV below and at $\sim2.5$~eV above $E_F$.

\begin{figure}
\includegraphics*[angle=270,width=0.39\textwidth]{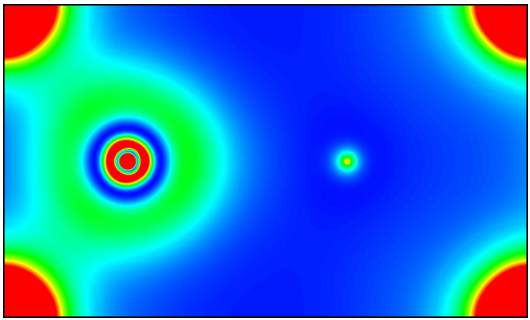}
\caption{Valence charge density (calculated using the FLAPW method) of
non-magnetic LiFeAs in the plane cutting through Fe, As, and Li nuclei. Fe
nuclei are at the corners of the plot. The linear color scale extends from 0
(blue) to 0.2 a.u.\ (red).} \label{CD}
\end{figure}

Fig.\ \ref{CD} shows the valence charge density in the vertical plane cutting
through the nearest Fe and As nuclei. The As sites are easily identified by the
presence of two nodes in the radial $4p$ wavefunction. Strong covalent bonds
between Fe and As are clearly seen. The filling fraction of the As $p$ states
is only 39\%, although the extended $p$ orbital spills out of the atomic sphere
(2.58 a.u.) somewhat. A fully developed covalent bond corresponds to a filling
fraction of 50\%. Note that the Pauling electronegativity difference between Fe
and As is only 0.35, so that an almost non-polar bond is expected.

For a strong covalent bond, the bonding-antibonding splitting is very sensitive
to the overlap integrals, and hence to the distance between the atoms
participating in the bond. Fig.\ \ref{bands}b shows the effect of moving As
atoms closer to Fe layers so that $R_{\mathrm{Fe-As}}$ is reduced from 2.42 to
2.33 \AA. This relatively small change in $R_{\mathrm{Fe-As}}$ has a large
effect on the band structure. An increase of the bonding-antibonding splitting
is evident; the antibonding Fe-As states move notably upward, pulling the Fe
partial density of states away from the Fermi level. The main change at the
Fermi level is the upward shift of the dispersive band near the Z point, which
contains a significant As $p_z$ weight. This upward shift of the antibonding
states results in the fourfold reduction of the local moment in the stripe
\cite{Yildirim-mag} phase (see Table I).

It is useful to view the Fe-As structure as being formed by two As semi-layers
adsorbed on each side of a free-standing Fe monolayer (which is compressed by
6\% compared to (001) layers in bulk bcc Fe). The strong hybridization of Fe
with As apparent in Figs.\ \ref{bands}a and \ref{CD} appears natural in view of
this analogy. In fact, the experimentally measured position of the (chemically
similar but slightly smaller) phosphorus adsorbed as a $c(2\times2)$ layer on
the Fe(001) surface \cite{Huff} translates to the Fe-P bond length of 2.27 \AA,
which is strikingly close to the Fe-As distance in Fe-As compounds.
\cite{NaCoO2}

Surfaces of magnetic transition metals often lose their magnetization under
adsorption of such elements as O, S, H, N, P, etc. First-principles
calculations for such surfaces often show reduced magnetic moments in the
surface layer which is strongly bound to the adsorbant;
\cite{Weinert,Chubb,Geng} this effect is also observed experimentally.
\cite{Landolt,Passek,Morrall} The magnetic ``dead layer'' on the surface
appears precisely due to the chemical bonding, which partially removes the
transition-metal $3d$ states from the Fermi level. Bonding and antibonding
surface subbands are common in such cases. \cite{Co-oxid} Even if the surface
retains some magnetic moment, it may be reduced compared to the bulk. Indeed,
as the material is magnetized and the Fermi level reaches the majority-spin
antibonding states, further exchange splitting becomes unfavorable. All these
arguments fully apply to Fe-As compounds. Although the surface of bcc Fe
usually remains magnetic under chemisorption, the local moment in the
``free-standing'' monolayer in Fe-As compounds is expected to be less stable,
because it is not supported by the magnetized bulk.

It is reasonable to hypothesize that the emptying of the antibonding states is
the primary factor responsible for the equilibrium position of As atoms above
the Fe layers. As soon as these states move above the Fermi level, further
reduction of $R_{\mathrm{Fe-As}}$ does not bring a large gain in the binding
energy. The proximity of the antibonding Fe-As states to the Fermi level
appears to be universal among the iron-pnictide compounds. The analysis
presented above indicates that covalent bonding competes with the tendency to
form local moments in the Fe layer. It is therefore quite natural that the Fe
magnetic moment is extremely sensitive to $R_{\mathrm{Fe-As}}$, as noted by
other authors.\cite{Yin,Mazin} Competition between covalency and magnetism can
also explain why band structure calculations with \emph{optimized} As positions
seem to be in better agreement with experiment compared to those that use
experimental As positions. \cite{Mazin} Indeed, while the local density
approximation or other exchange-correlation potentials used in density
functional theory (DFT) may err in the binding energies, it is reasonable to
expect that equilibrium As positions, whether in nature or in DFT, correspond
to a similar \emph{balance} between covalent binding and the tendency to form
local moments.

We now concentrate on the magnetic properties of LiFeAs. The so-called stripe
phase is believed to be the ground state for other iron-pnictide compounds;
\cite{Yildirim-mag,Mazin} we therefore focus on this state. The stabilization
energy at the experimental structural parameters is found to be 64 meV per Fe
site in FLAPW or 56 meV in LMTO-ASA; the local moment is 1.40 $\mu_B$ in FLAPW
or 1.49 $\mu_B$ in LMTO-ASA. Further analysis shows that the trends in the
exchange interaction in the stripe phase are insensitive to the particular
choice of ASA parameters; the following results are, therefore, quite robust.

Among the possible exchange mechanisms, superexchange and Fermi surface nesting
between electron and hole sheets are usually mentioned for Fe-As compounds.
\cite{Mazin,Yildirim-mag} We calculated the pair exchange parameters for LiFeAs
using the linear response technique \cite{Jij} with a subsequent division by
$\mathbf{S}_i\mathbf{S}_j$, where $\mathbf{S}_i$ is the total spin moment in
the atomic sphere at site $i$. These parameters map the energies of small
deviations from the reference state to the Heisenberg model $E=E_0-\sum_{ij}
J_{ij}S_{i}S_{j}$. They were analyzed as a function of $R_{\mathrm{Fe-As}}$,
doping level, and also of the fictitious external potential $V$ coupled to the
occupation $n_p$ of the As $p$ orbital (see below).

The calculated pair exchange parameters for several nearest neighbors in the
stripe phase are listed in Tables \ref{exch1} and \ref{exch2}. The Tables also
include the values $J_0=\sum_i p_{0i}J_{0i}$ where $p_{ij}=1$ or $-1$ for
parallel and antiparallel spin pairs, respectively; $J_0$ is proportional to
the Weiss field.

\begin{table*}[htb]
\caption{Pair exchange parameters in mRy as a function of $R_{\mathrm{Fe-As}}$.
$\Delta_z$ is the shift of the As layers toward Fe layers given as a fraction
of the lattice parameter $c$. Experimental structure corresponds to
$\Delta_z=0$. $M$ is the local moment in $\mu_B$. The exchange parameters
$J_\mathbf{R}$ are indexed by the crystallographic indices of the connecting
vector $\mathbf{R}$ in the simple tetragonal Fe sublattice, in which the $x$
axis is aligned parallel to the stripes. A spin pair is parallel if the second
index is even, and antiparallel otherwise.}
\begin{ruledtabular}
\begin{tabular}{lccccccccccccc}
$\Delta_z$ & $R_{\mathrm{Fe-As}}$ & $M$ & $J_{100}$ & $J_{010}$ & $J_{110}$ & $J_{200}$ & $J_{020}$ & $J_{120}$ & $J_{210}$ & $J_{001}$ & $J_{101}$ & $J_{011}$ & $J_0$\\
\hline
0       & 2.4204 &         1.28 &  -0.83  &  -2.61  &  -0.56  & -0.24  & -0.23  &  0.27  & -0.036  &  -0.063 &   -0.024  &  -0.035 & 6.7\\
$0.01c$ & 2.3972 &         1.05 &  -1.26  &  -2.96  &  -0.75  & -0.19  & -0.14  &  0.30  & -0.033  &  -0.053 &   -0.039  &  -0.049 & 8.1\\
$0.02c$ & 2.3739 &         0.81 &  -1.73  &  -3.11  &  -0.91  &  0.03  &  0.09  &  0.35  & -0.009 &  -0.044 &   -0.044  &  -0.057 & 9.8\\
$0.03c$ & 2.3517 &         0.57 &  -2.10  &  -3.10  &  -1.01  &  0.22  &  0.34  &  0.38  &  0.025  &  -0.032 &   -0.061  &  -0.052 & 11.6\\
$0.04c$ & 2.3294 &         0.31 &  -2.43  &  -2.90  &  -0.94  &  0.63  &  0.60  &  0.36  &  0.18  &  -0.032 &   -0.016  &  -0.028 & 13.6\\
\end{tabular}
\end{ruledtabular}
\label{exch1}
\end{table*}

\begin{table*}[htb]
\caption{Pair exchange parameters as a function of the external potential $V$
coupled to As $p$ occupation (see text). The last four lines are for $V=0$ with
the doping level per f.u.\ listed in the first column. The notation is the same
as in Table \ref{exch1}.}
\begin{ruledtabular}
\begin{tabular}{ccccccccccccc}
$V$ & $M$ & $J_{100}$ & $J_{010}$ & $J_{110}$ & $J_{200}$ & $J_{020}$ & $J_{120}$ & $J_{210}$ & $J_{001}$ & $J_{101}$ & $J_{011}$ & $J_0$\\
\hline
-1 eV  &   1.52 & -0.64  &  -2.68 &   -0.46  &  -0.20  &  -0.13  &  0.33  &  0.017  &  -0.087 &  -0.030  & -0.044 & 6.3\\
0      &   1.28 & -0.83  &  -2.61 &   -0.56  &  -0.24  &  -0.23  &  0.27  & -0.036  &  -0.063 &  -0.024  & -0.035 & 6.7\\
1 eV   &   1.10 & -0.70  &  -2.30 &   -0.65  &  -0.14  &  -0.22  &  0.24  & -0.068  &  -0.060 &  -0.037  & -0.052 & 7.5\\
2 eV   &   0.96 & -0.43  &  -1.86 &   -0.71  &  -0.019  &  -0.22  &  0.21  & -0.089  &  -0.058 &  -0.048  & -0.073 & 8.0\\
3 eV   &   0.84 & -0.06  &  -1.33 &   -0.78  &   0.051  &  -0.25  &  0.16  & -0.13  &  -0.040 &  -0.060  & -0.097 & 8.1\\
4 eV   &   0.70 &  0.30  &  -0.65 &   -0.87  &  -0.059  &  -0.32  &  0.033  & -0.25  &  -0.0082 & -0.068  & -0.122 & 7.7\\
5 eV   &   0.56 &  0.84  &   0.30 &   -0.79  &  -0.19  &  -0.27  & -0.087  & -0.33  &  -0.0013 & -0.121  & -0.154 & 6.6\\
\hline
$-0.2e$ & 1.20 &  -1.24  &  -1.26  &  -0.63  &  -0.12  &  -0.49  &  0.12 & -0.30  &    0.052  &  -0.028 &   -0.005 & 3.0\\
$-0.1e$ & 1.22 &  -1.19  &  -2.13  &  -0.54  &  -0.09  &  -0.29  &  0.34 & -0.19  &   -0.042  &  -0.034 &   -0.052 & 5.3\\
$+0.1e$ & 1.34 &  -0.56  &  -2.80  &  -0.71  &  -0.35  &  -0.38  &  0.15 & -0.02  &   -0.011  & -0.0042 &   -0.013 & 7.0\\
$+0.2e$ & 1.40 &  -0.41  &  -2.65  &  -0.88  &  -0.41  &  -0.49  & -0.038 & -0.01  &    0.003  & -0.022  &   -0.027 & 6.7\\

\end{tabular}
\end{ruledtabular}
\label{exch2}
\end{table*}

The exchange interaction is quite long-range; some parameters for pairs beyond
those included in the table are comparable with, say, $J_{020}$. The
contribution of the first two coordination spheres to $J_0$ declines steadily
from 0.86 at $\Delta_z=0$ to 0.35 at $\Delta_z=0.04c$ (Table \ref{exch1}). A
similar, but weaker trend is observed when $V$ is increased (Table
\ref{exch2}). Thus, the reduction of the local moment is accompanied by the
increase of the interaction range. Further, the ratio $J_0/J_{00}$, where
$J_{00}$ is defined similar to $J_{ij}$ and reflects the magnitude of the
on-site (Hund) exchange, steadily grows from 0.26 at $\Delta_z=0$ to 0.70 at
$\Delta_z=0.04c$. These trends indicate that the degree of itinerancy
significantly increases as $R_{\mathrm{Fe-As}}$ is decreased.

When the local moment is small, the exchange parameters are isotropic, as
required by symmetry. For larger moments there is a large anisotropy, in
particular for nearest neighbors (NN). For all Fe-As distances (Table
\ref{exch1}) the NN and next-nearest (NNN) exchange parameters are
antiferromagnetic (AFM); the exchange along the stripe is thus frustrated. The
ratio $J_{110}/J_{100}$ decreases from 0.68 at $\Delta_z=0$ (see Table
\ref{exch1}) to 0.39 at $\Delta_z=0.04c$; this trend tends to make the stripe
phase less stable.

We now discuss the dependence of exchange coupling on the external potential
$V$ coupled to the occupation of the As $p$ states. This is done by adding an
additional term $Vn_p$ to the Hamiltonian; DFT self-consistency is achieved for
each value of $V$. Adding $V$ results in the raising of both bonding and
antibonding Fe-As states, so that the antibonding states are moved away from
the Fermi level. \cite{screening} This change is expected to have a strong
effect on the magnitude of the superexchange interaction. \cite{superex} Fig.\
\ref{bands}c shows the energy bands obtained with $V=5$~eV. The band structure
close to the Fermi level is very similar to $V=0$ (Fig.\ \ref{bands}).
Therefore, direct exchange (including the ``nesting'' contribution) should be
insensitive to $V$. Thus, the analysis of trends associated with the As $p$
level shift will allow us to disentangle direct metallic exchange from
superexchange contributions.

Table \ref{exch2} shows that the increase of $V$ leads to the decrease and
eventual sign change of the NN coupling to ferromagnetic. The NNN exchange,
however, is almost unaffected. This result suggests that NN exchange has a
large AFM contribution from superexchange, while $J_{110}$ is primarily due to
direct metallic exchange including the nesting effects. Moreover, the direct
exchange contribution to the NN exchange appears to be ferromagnetic.

Table \ref{exch2} also shows the dependence of the exchange parameters on the
doping level, which was changed by varying the charge of the Li nuclei. The
local moment is quite insensitive to doping of $\pm20$\%, but a drastic change
of exchange parameters is apparent. Most notably, the anisotropy of NN exchange
interaction is reduced from $J_{100}/J_{010}\approx1$ at 20\% hole doping to
0.15 at 20\% electron doping. Also, the ratio $J_{110}/J_{100}$, which is
important for the stability of the stripe phase, increases from about 0.5 to
2.1. In the stripe phase the exchange splitting brings the Fermi level close to
the antibonding states. It is likely that the anisotropy of NN coupling
reflects the sensitivity of superexchange to the anisotropy of the electronic
structure. Interestingly, $J_{100}$ is much more sensitive to both the Fe-As
distance and doping than $J_{010}$.

The exchange parameters between the Fe layers are quite small and frustrated
for the stripe phase (the first three shown in the Tables are all AFM). These
parameters likely vary  between different Fe-As compounds.

In conclusion, layered iron-pnictide compounds are characterized by strong
covalent Fe-As bonding which is antagonistic to magnetism and responsible for
the huge magnetostructural coupling. The magnetic interaction is
long-range and shows increasing itinerancy with decreasing
$R_{\mathrm{Fe-As}}$. Superexchange makes a dominant AFM contribution to the NN
magnetic coupling, while the NNN coupling is mainly due to direct exchange. The
anisotropy of NN coupling is sensitive to doping.

\acknowledgments

We thank Mark van Schilfgaarde for the use of his LMTO codes and Igor Mazin for
useful comments. K.B. is a Cottrell Scholar of Research Corporation. Support
from the Nebraska Research Initiative is acknowledged. Work at Ames Laboratory
was supported by Department of Energy-Basic Energy Sciences, under Contract No.
DE-AC02-07CH11358.


\begin{thebibliography}{99}
\bibitem{NaCoO2} Interestingly, another superconductor Na$_x$CoO$_2$ also has widely
separated hexagonal Co monolayers with a monolayer of O internally
adsorbed on each side; the Co-O distance there is very close to that for the Co
surface with an adsorbed O monolayer.
\bibitem{screening} The self-consistent shift of the As $p$ states is much
smaller than $V$ due to electrostatic screening.
\end{thebibliography}
\end{document}